 \documentclass[final,1p,times]{elsarticle}

 \usepackage{graphicx}

\usepackage{amssymb}
\usepackage{amsmath}

\journal{Journal of Physics and Chemistry of Solids}

\begin{document}
\begin{frontmatter}
\title{Nuclear Magnetic Relaxation and Knight Shift Due to Orbital Interaction 
in Dirac Electron Systems}

\author[mymainaddress]{Hideaki Maebashi
\corref{mycorrespondingauthor}
}
\cortext[mycorrespondingauthor]{Corresponding author}
\author[mymainaddress]{Tomoki Hirosawa}
\author[mymainaddress]{Masao Ogata}
\author[mysecondaryaddress]{Hidetoshi Fukuyama}

\address[mymainaddress]{Department of Physics, University of Tokyo, Bunkyo, Tokyo 113-0033, Japan}
\address[mysecondaryaddress]{Department of Applied Physics, Tokyo University of Science, Shinjuku, Tokyo 162-8601, Japan}

\begin{abstract}
We study the nuclear magnetic relaxation rate and Knight shift in the presence of the orbital and quadrupole interactions for three-dimensional Dirac electron systems (e.g., bismuth--antimony alloys). 
By using recent results of the dynamic magnetic susceptibility and permittivity, we obtain rigorous results of the relaxation rates $(1/T_1)_{\rm orb}$ and $(1/T_1)_{\rm Q}$, which are due to the orbital and quadrupole interactions, respectively, and show that $(1/T_1)_{\rm Q}$ gives a negligible contribution compared with $(1/T_1)_{\rm orb}$. 
It is found that $(1/T_1)_{\rm orb}$ exhibits anomalous dependences on temperature $T$ and chemical potential $\mu$.
When $\mu$ is inside the band gap, $(1/T_1)_{\rm orb} \sim T ^3 \log (2 T/\omega_0)$ for temperatures above the band gap, where $\omega_0$ is the nuclear Larmor frequency. 
When $\mu$ lies in the conduction or valence bands, $(1/T_1)_{\rm orb} \propto T k_{\rm F}^2 \log (2 |v_{\rm F}| k_{\rm F}/\omega_0)$ for low temperatures, 
where $k_{\rm F}$ and $v_{\rm F}$ are the Fermi momentum and Fermi velocity, respectively. 
The Knight shift $K_{\rm orb}$ due to the orbital interaction also shows anomalous dependences on $T$ and $\mu$.   
It is shown that $K_{\rm orb}$ is negative and its magnitude significantly increases with decreasing temperature when $\mu$ is located in the band gap. 
Because the anomalous dependences in $K_{\rm orb}$ 
is caused by the interband particle-hole excitations across the small band gap  
while $\left( 1/T_1 \right)_{\rm orb}$ is governed by the intraband excitations, 
the Korringa relation does not hold in the Dirac electron systems.
\end{abstract}


\end{frontmatter}

\section{Introduction}
\label{S:1}

Bismuth is a narrow-gap material with strong spin--orbit coupling 
and its low-energy properties are described by Dirac 
electrons~\cite{Wolff64,Fuseya15}. 
One of the characteristic properties in bismuth is its large diamagnetism 
which has been known since the18th century. 
More importantly, the diamgnetism of bismuth--antimony alloys Bi$_{1-x}$Sb$_{x}$ 
significantly increases with decreasing temperature
in the band insulator regime of $0.07 \lesssim x \lesssim 0.2$~\cite{Shoenberg36,Wehrli68}.
This behavior is distinct from both the core diamagnetism of atoms and 
Landau diamagnetism in metals. 
The permittivity was also found large in bismuth~\cite{Boyle60,Edelman75}, 
which turned out to be related to the large diamagnetism~\cite{Maebashi17}. 
Recently, the $\beta$-NMR measurement in $\mathrm{Bi}_{0.9}\mathrm{Sb}_{0.1}$ 
showed anomalous temperature dependence in 
the nuclear magnetic relaxation time $T_1$~\cite{MacFarlane14}, 
attracting a renewed interest in relaxation mechanism due to Dirac and Weyl electron 
systems~\cite{Okvatovity16,Yasuoka16,Hirosawa17}. 

Based on the Wolff Hamiltonian~\cite{Wolff64}, 
which is derived by applying the $k \cdot p$ theory 
to a narrow-gap material with strong spin--orbit coupling~\cite{Luttinger55,Cohen60}, 
the large diamagnetism in bismuth has been theoretically explained by an interband effect of the magnetic field~\cite{Fukuyama70}.
This led to the construction of a general theory of orbital magnetism~\cite{Fukuyama71} 
followed by recent progress including its extension in spin--orbit coupled systems~\cite{Ogata15,Ogata16-1,Ogata16-2,Ogata17}. 
More recently, it was pointed out that diamagnetism and an enhancement in the permittivity are directly linked to each other because of effective Lorentz covariance in the Dirac Hamiltonian, 
which is essentially identical to the Wolff Hamiltonian~\cite{Maebashi17}.   
Thus, the interband effects of an electromagnetic field induce 
not only large diamagnetism but also a significant enhancement in the permittivity, 
and this is a general property of Dirac electron systems. 

As we discussed above, the orbital magnetism plays an important role in 
a narrow-gap material with strong spin--orbit coupling. 
The contribution of orbital magnetism could be further discussed in the nuclear spin relaxation.
In general, nuclear magnetic relaxation is caused by magnetic and quadrupole interactions between a nuclear magnetic moment and surrounding 
electrons~\cite{Abragam61,Obata63,Obata64}. 
The magnetic interaction consists of the Fermi contact, dipole, and orbital interactions. 
Among these, the orbital interaction gives rise to anomalous dependence 
of the nuclear magnetic relaxation time $T_1$ on temperature $T$.
In Weyl fermion systems, a recent theory shows that 
$1/T_1T \propto E^2 \log (E/\omega_0)$ due to the orbital interaction  
where $E$ is the maximum of temperature and chemical potential, 
and $\omega_0$ is the nuclear Larmor frequency~\cite{Okvatovity16}. 
This is consistent with the experimental observation of the NQR measurement in TaP~\cite{Yasuoka16}. 
More recently, the $T$ dependence of $1/T_1$ due to the orbital interaction has 
theoretically been obtained for Dirac electron systems~\cite{Hirosawa17}. 
The obtained result is a little more complicated because 
of the existence of a gap, and 
partly explains the experimental observation of 
the $\beta$-NMR measurement in $\mathrm{Bi}_{0.9}\mathrm{Sb}_{0.1}$~\cite{MacFarlane14}. 

In this paper, by using the results of the dynamic magnetic susceptibility and permittivity  
in Ref.~\cite{Maebashi17},  
we systematically derive the $T$ dependences of the nuclear magnetic relaxation rates 
$(1/T_1)_{\rm orb}$ and $(1/T_1)_{\rm Q}$, which are due to the orbital and quadrupole interactions, respectively, in three-dimensional (3D) Dirac electron systems. 
The result of $(1/T_1)_{\rm orb}$  
has already been published in Ref.~\cite{Hirosawa17}, but some errors there are corrected 
in this paper. 
We present a rigorous result of $(1/T_1)_{\rm orb}$ for the 3D Dirac electrons, 
which correctly reproduces the two limiting cases of the free-electron gas and Weyl fermions, 
and give a prediction on $(1/T_1)_{\rm orb}$ for quasi-2D Dirac electrons. 
We also discuss the $T$ dependence of the uniform and static orbital magnetic susceptibility 
and the Knight shift in Dirac electron systems.
Throughout the paper, we take $\hbar = k_{\rm B} = 1$ for simplicity.

\section{Diamagnetism of Dirac electrons}
\label{S:2}

Our Hamiltonian is described by the 3D Dirac Hamiltonian as
\begin{align}
\label{eq:Dirac}
H = \sum_{\boldsymbol k} \psi_{\boldsymbol k}^{\dagger} \gamma^0
\left[ {c^*}  {\boldsymbol k} \, {\bf \cdot} \, 
{\boldsymbol \gamma} + \Delta  \right] \psi_{\boldsymbol k} ,
\end{align}
where
$\psi_{\boldsymbol k}$ ($\psi_{\boldsymbol k}^{\dagger}$) 
corresponds to the annihilation (creation) operator of the conduction and valence band 
electrons with spin degeneracy,
$\gamma^0$ and ${\boldsymbol \gamma} = (\gamma^1, \gamma^2, \gamma^3)$ 
are the gamma matrices, $\Delta$ is a half band gap, and 
${c^*} \equiv \sqrt{\Delta/{m^*}}$ with ${m^*}$ as the effective electron mass. 
In this paper, we do not consider anisotropy of the effective electron mass, 
which is taken into account in the Wolff Hamiltonian~\cite{Wolff64}.

In this section, we consider the uniform and static orbital magnetic susceptibility
$\chi_{\rm m} = \chi_{\rm m} (T, \mu)$ as a function of temperature $T$ and 
chemical potential $\mu$. 
The finite-temperature susceptibility can be expressed as an integral of the 
zero-temperature susceptibility $\chi_{\rm m} (0, \epsilon)$ 
with respect to $\epsilon$ as~\cite{Maebashi17,Maldague78}
\begin{align}
\chi_{\rm m} (T, \mu)
&= 
\int_{-\infty}^{\infty} d \epsilon 
\left[ - \frac{\partial f (\epsilon)}{\partial \epsilon} \right]
\chi_{\rm m} (0, \epsilon) ,
\label{eq:diamag1}
\end{align}
where $f(\epsilon) = [ e^{(\epsilon - \mu)/T} + 1 ]^{-1}$ is the Fermi distribution function. 
The zero-temperature susceptibility for Dirac electrons has been previously obtained 
as~\cite{Fukuyama70,Fukuyama71,Fuseya14}
\begin{align}
\chi_{\rm m} (0, \mu) &= - \frac{2 \alpha}{3 \pi} \frac{{c^*}}{c}\times
\left\{ 
\begin{array}{lr}
\displaystyle{
\log \frac{E_{\Lambda}}{|\mu| +\!\! \sqrt{\mu^2 - \Delta^2}},
}
&
|\mu| > \Delta
\\
&
\\
\displaystyle{
\log \frac{E_{\Lambda}}{\Delta} ,
}
&
|\mu| \leq \Delta
\end{array}
\right. ,
\label{eq:diamag2}
\end{align}
where $\alpha = \frac{e^2}{4 \pi \varepsilon_0 c} \approx \frac{1}{137}$ is the 
fine-structure constant and $E_{\Lambda}$ is a bandwidth cutoff. 
By substituting Eq.~(\ref{eq:diamag2}) into Eq.~(\ref{eq:diamag1}), 
we can calculate the temperature dependence of the susceptibility.
\begin{figure}[h]
\begin{center}
\includegraphics[scale=0.38,keepaspectratio]{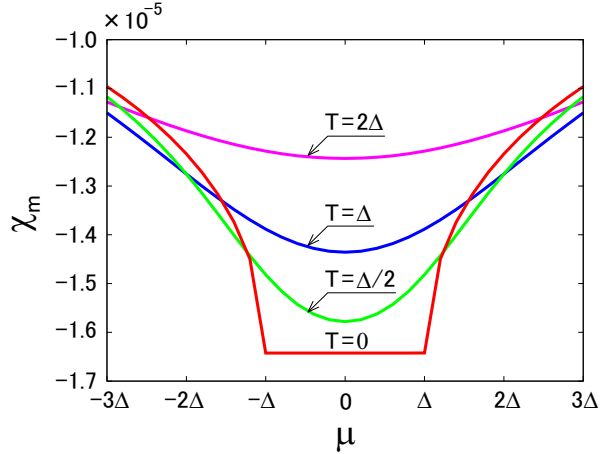}
\end{center}
\caption[Fig.1]{(Color online) Plots of the orbital magnetic susceptibility $\chi_{\rm m}$ 
as functions of the chemical potential $\mu$ 
for temperatures of $T = 2 \Delta$, $T = \Delta$, $T = \Delta/2$, and $T=0$ 
with ${c^*}/c = 2.0 \times 10^{-3}$ and $E_{\Lambda}/\Delta = 200$. } 
\label{fig:1}
\end{figure}

In Fig.~\ref{fig:1}, we plot $\chi_{\rm m}$ thus calculated 
as functions of $\mu$ for several temperatures with ${c^*}/c = 2.0 \times 10^{-3}$ and $E_{\Lambda}/\Delta = 200$. 
Note that its anomalous dependences on $T$ and $\mu$ are due to an interband effect of the magnetic field as emphasized in Sect. \ref{S:1}.
When the temperature is comparable with the band gap $2\Delta$, 
$\chi_{\rm m}$ has only a weak dependence on $\mu$. 
This situation is realized about at room temperature for 
bismuth--antimony alloys Bi$_{1-x}$Sb$_{x}$ since $\Delta \sim 10$ meV~\cite{Wehrli68}.
With decreasing temperature, however, the magnitude of $\chi_{\rm m}$ 
significantly increases in the band insulator regime of $|\mu| \lesssim \Delta$. 
This behavior is consistent with the temperature dependence of $\chi_{\rm m}$ 
observed experimentally in bismuth--antimony alloys~\cite{Wehrli68}. 
It is to be noted that when the effect of effective-mass anisotropy of actual material 
Bi$_{1-x}$Sb$_{x}$ is taken into account,  
the orbital magnetic susceptibility becomes about ten times larger in the direction
perpendicular to the trigonal axis~\cite{Fukuyama70}. 
This fact ensures a good agreement between theory and experiment.

\section{Dynamical Correlation Functions}
\label{S:3}

For the Dirac Hamiltonian, Eq.~(\ref{eq:Dirac}), 
the electric current and electric charge-density operators are given by 
$- e {\boldsymbol j}_{\boldsymbol q} = 
- e {c^*} \sum_{\boldsymbol k} 
\psi_{{\boldsymbol k}+{\boldsymbol q}}^{\dagger} \gamma^0 {\boldsymbol \gamma}\, \psi_{{\boldsymbol k}}$ 
and $- e \rho_{\boldsymbol q} = 
-e \sum_{\boldsymbol k} 
\psi_{{\boldsymbol k}+{\boldsymbol q}}^{\dagger} \psi_{{\boldsymbol k}}$, 
respectively, where $e$ ($>0$) is the elementary charge. 
In this section, we derive useful equations for the dynamical correlation functions 
associated with these operators.

The fluctuation--dissipation theorem tells us that 
the current-current correlation function is related to the real part of 
the conductivity tensor $\sigma_{ij\,} (q, \omega)$ as~\cite{Kubo57}
\begin{align}
\label{eq:current-current}
e^2  \! \int_{-\infty}^{\infty} dt \, \langle\,  
j_{\boldsymbol q}^{\,i} (t) \, j_{-{\boldsymbol q}}^{\, j} \,\rangle \, e^{i \omega t}
&= \frac{2 \omega}{1- e^{- \omega/T}} {\rm Re} \sigma_{ij\,} (q, \omega) 
\end{align}
for $i,j = x, y, z$ and where $j_{\boldsymbol q}^{\, i} (t) = e^{i (H - \mu N) t} j_{\boldsymbol q}^{\, i} e^{- i (H - \mu N) t}$ and $\langle\, \cdots \rangle$ denotes the grand canonical average for 
$H - \mu N$ with $N=\sum_{\boldsymbol k} 
\psi_{{\boldsymbol k}}^{\dagger} \psi_{{\boldsymbol k}}$ as the number operator. 
We can separate $\sigma_{ij\,} (q, \omega)$ into the transverse and longitudinal components as
\begin{align}
\label{eq:conductivity}
\sigma_{ij\,} (q, \omega) &= 
\left( \delta_{ij} - \frac{q_i q_j}{q^2} \right) \sigma_{\rm T} (q, \omega) 
+ \frac{q_i q_j}{q^2} \sigma_{\rm L} (q, \omega).
\end{align}
Then, Maxwell's equations in matter lead to the fact that the transverse conductivity 
$\sigma_{\rm T} (q, \omega)$ and longitudinal conductivity $\sigma_{\rm L} (q, \omega)$ 
can be expressed in terms of the dynamic magnetic susceptibility $\chi_{\rm m} (q, \omega)$ and relative permittivity $\varepsilon_{\rm r} (q, \omega)$ as~\cite{Maebashi17,Jackson99}
\begin{align}
\label{eq:transverse}
\sigma_{\rm T} (q, \omega) &= \frac{q^2}{i \omega \mu_0} 
\left(\, \chi_{\rm m} (q, \omega) + \frac{\omega^2}{c^2 q^2}  
\left[ 
\varepsilon_{\rm r} (q, \omega) - 1
\right]  \,\right) ,
\\
\label{eq:longitudinal}
\sigma_{\rm L} (q, \omega) &= - i \omega \varepsilon_0  
\left[ 
\varepsilon_{\rm r} (q, \omega) - 1
\right] .
\end{align}
The first and second terms in Eq.~(\ref{eq:transverse}) 
correspond to the magnetization and polarization currents, respectively.
The uniform and static orbital magnetic susceptibility $\chi_{\rm m}$ in Sect.~\ref{S:1} 
is given by 
the so-called $q$ limit of $\chi_{\rm m}(q, \omega)$, 
i.e., $\lim_{q \to 0}\chi_{\rm m}(q, 0)$. 

On the other hand, the fluctuation--dissipation theorem for 
the charge-charge correlation function leads to 
\begin{align}
\label{eq:charge-charge}
e^2  \! \int_{-\infty}^{\infty} dt \, \langle\,  
\rho_{\boldsymbol q} (t) \, \rho_{-{\boldsymbol q}} \,\rangle \, e^{i \omega t}
&= \frac{2 \varepsilon_0 q^2}{1- e^{- \omega/T}} {\rm Im} \varepsilon_{\rm r} (q, \omega) ,
\end{align}
where 
$\rho_{\boldsymbol q} (t) = e^{i (H - \mu N) t} \rho_{\boldsymbol q} e^{- i (H - \mu N) t}$. 
From Eqs.~(\ref{eq:longitudinal}) and~(\ref{eq:charge-charge}), 
we see that the charge-charge correlation function can be described only from the longitudinal conductivity.

By use of the imaginary parts of $\chi_{\rm m}(q, \omega)$ 
and $\varepsilon_{\rm r}(q, \omega)$ in 
Ref.~\cite{Maebashi17}, we can calculate the above dynamical correlation functions explicitly.
The expressions for ${\rm Im} \chi_{\rm m} (q, \omega)$ and 
${\rm Im} \varepsilon_{\rm r} (q, \omega)$ that is suitable for our purpose 
has the following form (see Supplemental Material of Ref.~\cite{Maebashi17}):
\begin{align}
{\rm Im} \chi_{\rm m} (q, \omega)
&= \frac{e^2 \mu_0 {c^*}^4}{8 \pi^2} \!\!
\int_{-\infty}^{\infty}  \! d \epsilon \, 
\left( 1+ \frac{4 (\epsilon^2 - \Delta^2) + \omega^2 }{{c^*}^2 \! q^2} 
- \frac{12 \epsilon^2 \omega^2}{{c^*}^4 \! q^4} \right)
\left[ f \left( \epsilon_- \right) 
\!-\!  f \left( \epsilon_+  \right)  \right] 
\nonumber
\\
&\quad \times
{\rm sgn} \! \left( \epsilon_- \right) 
{\rm sgn} \! \left( \epsilon_+ \right) 
\int \! d^3 k \,\delta \left( \epsilon_-^2 - E_k^2 \right) 
\int \! d^3 k' \,\delta \left( \epsilon_+^2 - E_{k'}^2 \right)
\delta^3 \! \left( {\boldsymbol k} - {\boldsymbol k}' + {\boldsymbol q} \right) ,
\label{eq:main1}
\\
{\rm Im} \varepsilon_{\rm r} (q, \omega)
&= \frac{e^2 \mu_0 {c^*}^4 }{4 \pi^2} \left( \frac{c}{{c^*}} \right)^2
\int_{-\infty}^{\infty}  \! d \epsilon \,
\left(\frac{4 \epsilon^2}{{c^*}^2 \! q^2} - 1 \right) 
\left[ f \left( \epsilon_- \right) 
\!-\!  f \left( \epsilon_+  \right)  \right]
\nonumber
\\
&\quad \times
{\rm sgn} \! \left( \epsilon_- \right) 
{\rm sgn} \! \left( \epsilon_+ \right) 
\int \! d^3 k \,\delta \left( \epsilon_-^2 - E_k^2 \right) 
\int \! d^3 k' \,\delta \left( \epsilon_+^2 - E_{k'}^2 \right)
\delta^3 \! \left( {\boldsymbol k} - {\boldsymbol k}' + {\boldsymbol q} \right) , 
\label{eq:main2}
\end{align}
where $\epsilon_\pm = \epsilon \pm \omega/2$ and $E_k = \sqrt{{c^*}^2 k^2 + \Delta^2}$ 
is the dispersion of Dirac electrons.

\section{Nuclear Magnetic Resonance}
\label{S:4}

\subsection{Interaction Hamiltonian}

Interaction Hamiltonian $H'$ between a nuclear magnetic moment and surrounding electrons 
can be separated into a magnetic term $H'_{\rm M}$ and a quadrupole term $H'_{\rm Q}$. 
The latter exists only for $I > 1/2$, where
$I$ is the quantum number of the nuclear magnetic moment~\cite{Abragam61}. 
In this paper, we consider $H'_{\rm Q}$ and the orbital interaction 
$H'_{\rm orb}$ for the magnetic interaction $H'_{\rm M}$. 
Then, our interaction Hamiltonian $H'$ can be written in the form of gauge coupling as
\begin{align}
H' = H'_{\rm orb} + H'_{\rm Q}
\end{align}
with
\begin{align}
H'_{\rm orb} &= e \sum_{\boldsymbol q} {\boldsymbol j}_{\boldsymbol q} {\bf \cdot}\, 
{\boldsymbol a}_{-{\boldsymbol q}} ,
\label{eq:Horb}
\\
H'_{\rm Q} &= - e \sum_{\boldsymbol q} \rho_{\boldsymbol q} \phi_{-{\boldsymbol q}} .
\label{eq:HQ}
\end{align}
The vector potential ${\boldsymbol a}_{\boldsymbol q}$ produced by a nuclear magnetic dipole 
is given by~\cite{Abragam61,Jackson99}
\begin{align}
\label{eq:orbital}
{\boldsymbol a}_{\boldsymbol q} = - i \mu_0 \gamma_{\rm n}  {\boldsymbol I}
\,{\bf \times}\, \frac{{\boldsymbol q}}{q^2} ,
\end{align}
where $\gamma_{\rm n}$ is the gyromagnetic ratio of a nucleus and 
${\boldsymbol I} = (I_x, I_y, I_z)$ is a nuclear angular momentum operator. 
The scalar potential $\phi_{\boldsymbol q}$ produced by a nuclear electric quadrupole 
is given by~\cite{Abragam61,Jackson99}
\begin{align}
\label{eq:quadrupole}
\phi_{\boldsymbol q} = - \frac{e Q}{6 \varepsilon_0 I (2I - 1)} \sum_{{i}, {j}}
\left[ \frac{3}{2} \left( I_{{i}} I_{{j}} + I_{{j}} I_{{i}} \right) 
- \delta_{{i}{j}} {\boldsymbol I}^2 \right] \frac{q_{{i}} q_{{j}}}{q^2} ,
\end{align}
where $Q$ is the nuclear quadrupole moment and ${i}, {j} = x, y, z$. 

Because ${\boldsymbol j}_{\boldsymbol q} {\bf \cdot}\, 
{\boldsymbol a}_{-{\boldsymbol q}}$ can be written as
${\boldsymbol j}_{\boldsymbol q} {\bf \cdot}\, 
{\boldsymbol a}_{-{\boldsymbol q}} = 
i \mu_0 \gamma_{\rm n}  {\boldsymbol I}
\,{\bf \cdot}\, 
({\boldsymbol q} 
\,{\bf \times}\, {\boldsymbol j}_{\boldsymbol q})/q^2$, 
$H'_{\rm orb}$ can also been written as
$H'_{\rm orb} = - \gamma_{\rm n}  {\boldsymbol I}
\,{\bf \cdot}\, {\boldsymbol B}_{\rm eff}$, 
where an effective magnetic induction field is given by
\begin{align}
\label{Beff}
{\boldsymbol B}_{\rm eff} &= - i e \mu_0 \sum_{\boldsymbol q} 
\frac{{\boldsymbol q} \,{\bf \times}\, {\boldsymbol j}_{\boldsymbol q}}{q^2} .
\end{align}
Thus, the orbital interaction couples only to the transverse component of 
current~\cite{Lee91,Chklovskii92}. 
On the other hand, the continuity equation 
$- i \omega \rho_{\boldsymbol q} = - i {\boldsymbol q} \,{\bf \cdot}\, {\boldsymbol j}_{\boldsymbol q}$ leads to 
$\rho_{\boldsymbol q} \phi_{-{\boldsymbol q}} = \phi_{-{\boldsymbol q}} 
({\boldsymbol q} \,{\bf \cdot}\, {\boldsymbol j}_{\boldsymbol q})/\omega$ in $H'_{\rm Q}$, 
so that the quadrupole interaction couples only to 
the longitudinal component of current.

\subsection{Knight shift}
\label{S:4.2}

In an external magnetic induction field ${\boldsymbol B}_0 = (0,0,B_0)$ 
oriented along the $z$ axis, 
the energy $E_m$ of a nuclear spin state $|m \rangle$   
is given by $E_m = - m \gamma_{\rm n} B_0$, where $m$ is a quantum number of $I_z$. 
Because our electron systems are isotropic, 
$H'_{\rm Q}$ does not lead to any shift in $E_m$. 
However, $H'_{\rm orb}$ gives rise to a shift as 
$\Delta E_m = - m \gamma_{\rm n}  K_{\rm orb} B_0$, 
which is the expectation value of  
$H'_{\rm orb} = - \gamma_{\rm n}  {\boldsymbol I}\,{\bf \cdot}\, {\boldsymbol B}_{\rm eff}$. 
Here the Knight shift $K_{\rm orb}$ due to the orbital interaction is 
defined through $\langle {\boldsymbol B}_{\rm eff} \rangle = K_{\rm orb} {\boldsymbol B}_0$.
From Eq.~(\ref{Beff}), for a static external field ${\boldsymbol B}_{\boldsymbol q}$, 
the thermodynamic average $\langle {\boldsymbol B}_{\rm eff} \rangle$ of the effective field 
${\boldsymbol B}_{\rm eff}$ can be written as
\begin{align}
\langle {\boldsymbol B}_{\rm eff} \rangle 
&= - i e \mu_0 \sum_{\boldsymbol q} 
\frac{{\boldsymbol q} \,{\bf \times}\, \delta {\boldsymbol j}_{\boldsymbol q}}{q^2} ,
\end{align}
where $-e \delta {\boldsymbol j}_{\boldsymbol q} = i {\boldsymbol q} \,{\bf \times} 
{\boldsymbol M}_{\boldsymbol q}$ with a magnetization 
${\boldsymbol M}_{\boldsymbol q} = \mu_0^{-1} \chi_{\rm m}(q,0) 
{\boldsymbol B}_{\boldsymbol q}$. 
Thus, we obtain 
$\langle {\boldsymbol B}_{\rm eff} \rangle = \sum_{\boldsymbol q} \chi_{\rm m}(q,0) 
{\boldsymbol B}_{\boldsymbol q}$. 
Because ${\boldsymbol B}_{\boldsymbol q} \approx {\boldsymbol B}_0 \delta_{{\boldsymbol q}, {\bf 0}}$ for an almost uniform magnetic induction field ${\boldsymbol B}_0$, 
$\langle {\boldsymbol B}_{\rm eff} \rangle = \chi_{\rm m}{\boldsymbol B}_0$ 
and $K_{\rm orb}$ is just given by the orbital magnetic susceptibility $\chi_{\rm m}$ 
studied in Sect.~\ref{S:2}.

\subsection{Relaxation rate}

The nuclear magnetic relaxation time $T_1$ is given by~\cite{Abragam61}
\begin{align}
\frac{1}{T_1} = \frac{1}{2} \frac{\sum_{m,m'} W_{m \to m'} (E_m -E_{m'})^2}{\sum_{m} E_m^2},
\label{eq:T1total}
\end{align}
where $W_{m \to m'}$ is a transition probability from one nuclear spin state $|m\rangle$ to 
another nuclear spin state $|m'\rangle$. 
From Eq.~(\ref{eq:orbital}), 
the orbital interaction gives rise to $m \to m \pm 1$ transitions, 
whose probability is denoted as $W_{m \to m \pm 1}^{\rm orb}$. 
From Eq.~(\ref{eq:quadrupole}), on the other hand,
the quadrupole interaction leads to both $m \to m \pm 1$ and 
$m \to m \pm 2$ transitions, 
whose probabilities are denoted as $W_{m \to m \pm 1}^{\rm Q1}$ and 
$W_{m \to m \pm 2}^{\rm Q2}$, respectively.

The above transition probabilities can be derived from 
Fermi's golden rule and then related to the dynamical correlation functions 
studied in Sect.~\ref{S:3}.
From the fluctuation--dissipation theorem, Eqs.~(\ref{eq:current-current}) and 
(\ref{eq:charge-charge}), and the simple forms of Eqs.~(\ref{eq:Horb})--(\ref{eq:orbital}), 
we obtain
\begin{align}
W_{m \to m \pm 1}^{\rm orb} 
&= 2 T 
\sum_{\boldsymbol q} 
{\rm Re} \sigma_{\rm T} (q, \omega_0)
\left|\langle m \pm 1|{\boldsymbol a}_{{\boldsymbol q}}|m\rangle\right|^2 ,
\label{eq:16}
\\
W_{m \to m \pm 1}^{\rm Q1} 
&= 2 \varepsilon_0 T \!
\sum_{\boldsymbol q} 
q^2 \frac{{\rm Im} \varepsilon_{\rm r} (q, \omega_0)}{\omega_0}
\left|\langle m \pm 1|\phi_{\boldsymbol q}|m\rangle\right|^2 ,
\label{eq:17}
\\
W_{m \to m \pm 2}^{\rm Q2} 
&= 2 \varepsilon_0 T \!
\sum_{\boldsymbol q} 
q^2 \frac{{\rm Im} \varepsilon_{\rm r} (q, 2\omega_0)}{2 \omega_0}
\left|\langle m \pm 2|\phi_{\boldsymbol q}|m\rangle\right|^2 ,
\label{eq:18}
\end{align}
where $\omega_0 = \gamma_{\rm n} B_0 \ll T$ is the nuclear Larmor frequency. 
From Eqs.~(\ref{eq:orbital}) and ~(\ref{eq:quadrupole}), 
the vector and scalar potential parts in Eqs.~(\ref{eq:16})--(\ref{eq:18}) are calculated as
\begin{align}
\left|\langle m \pm 1|{\boldsymbol a}_{{\boldsymbol q}}|m\rangle\right|^2
&= \left(\frac{\mu_0 \gamma_{\rm n}}{2} \right)^2
\frac{q^2 + q_z^2}{q^4}
\left| \langle m \pm 1|I_{\pm}|m\rangle \right|^2 ,
\\
\left|\langle m \pm 1|\phi_{\boldsymbol q}|m\rangle\right|^2 
&= \left( \frac{e Q}{4 \varepsilon_0 I (2I - 1)}  \right)^2
\frac{q_z^2 \left( q_x^2 + q_y^2 \right)}{q^4}
\left| \langle m \pm 1|I_{\pm} I_z + I_z I_{\pm}|m\rangle \right|^2 ,
\\
\left|\langle m \pm 2|\phi_{\boldsymbol q}|m\rangle\right|^2 
&= \left( \frac{e Q}{4 \varepsilon_0 I (2I - 1)}  \right)^2
\frac{\left( q_x^2 - q_y^2 \right)^2 + q_x^2 q_y^2}{4 q^4}
| \langle m \pm 2|I_{\pm}^2|m\rangle |^2 ,
\end{align}
where $I_{\pm} = I_x \pm i I_y$. 
Then the transition probabilities can be written as
\begin{align}
W_{m \to m \pm 1}^{\rm orb} &= W_{\rm orb} | \langle m \pm 1|I_{\pm}|m\rangle |^2,
\label{eq:separate1}
\\
W_{m \to m \pm 1}^{\rm Q1} &= W_{\rm Q1} | \langle m \pm 1|I_{\pm} I_z + I_z I_{\pm}|m\rangle |^2, 
\label{eq:separate2}
\\
W_{m \to m \pm 2}^{\rm Q2} &= W_{\rm Q2} | \langle m \pm 2|I_{\pm}^2|m\rangle |^2,
\label{eq:separate3}
\end{align}
where
\begin{align}
W_{\rm orb}
&= \frac{2}{3} \mu_0^2 \gamma_{\rm n}^2 T 
\int \frac{d^3q}{(2 \pi)^3} \frac{1}{q^2}
{\rm Re} \sigma_{\rm T} (q, \omega_0)
\nonumber
\\
&= \frac{2}{3} \mu_0 \gamma_{\rm n}^2 T
\int \frac{d^3q}{(2 \pi)^3}  
\frac{ {\rm Im} 
\left[ \chi_{\rm m} (q, \omega_0) + \frac{\omega_0^2}{c^2 q^2} \varepsilon_{\rm r} (q, \omega_0) 
\right]}{\omega_0} ,
\label{eq:Worb}
\\
W_{\rm Q1} 
&= 
\frac{T}{60 \varepsilon_0} \left( \frac{e Q}{I (2I - 1)}  \right)^2
\int \frac{d^3q}{(2 \pi)^3} 
q^2 \frac{{\rm Im} \varepsilon_{\rm r} (q, \omega_0)}{\omega_0} ,
\label{eq:WQ1}
\\
W_{\rm Q2} 
&= 
\frac{T}{96 \varepsilon_0} \left( \frac{e Q}{I (2I - 1)}  \right)^2
\int \frac{d^3q}{(2 \pi)^3} 
q^2 \frac{{\rm Im} \varepsilon_{\rm r} (q, 2\omega_0)}{2 \omega_0} .
\label{eq:WQ2}
\end{align}
In Eq.~(\ref{eq:Worb}), we have used Eq.~(\ref{eq:transverse}).

Substitution of Eqs.~(\ref{eq:separate1})--(\ref{eq:separate3}) into Eq.~(\ref{eq:T1total}) 
and straightforward algebra lead to the fact that  
$1/T_1$ is given by the sum of the relaxation rate $\left( 1/T_1 \right)_{\rm orb}$ due to the orbital interaction and the relaxation rate $\left( 1/T_1 \right)_{\rm Q}$ due to the quadrupole interaction, where 
$\left( 1/T_1 \right)_{\rm orb} = 2 W_{\rm orb}$ 
and 
$\left( 1/T_1 \right)_{\rm Q} 
= (2/5) (2I-1)(2I+3) (W_{\rm Q1}+4W_{\rm Q2})$~\cite{Abragam61,Obata64}. 
Then we obtain $\left( 1/T_1 \right)_{\rm orb}$ and $\left( 1/T_1 \right)_{\rm Q}$ 
in terms of the dynamic magnetic susceptibility $\chi_{\rm m} (q, \omega)$ 
and relative permittivity $\varepsilon_{\rm r} (q, \omega)$ as
\begin{align}
\left(\frac{1}{T_1}\right)_{\rm orb}
&= \frac{4}{3} \mu_0 \gamma_{\rm n}^2 \frac{T}{\omega_0}
\int \frac{d^3q}{(2 \pi)^3}  
\left[ {\rm Im} \chi_{\rm m} (q, \omega_0) 
+ \frac{\omega_0^2}{c^2 q^2} {\rm Im}  \varepsilon_{\rm r} (q, \omega_0) 
\right],
\label{eq:general1}
\\
\left(\frac{1}{T_1}\right)_{\rm Q}
&= 
\frac{7 (2 I + 3)}{300\, I^2 (2I - 1)} 
\frac{e^2 Q^2}{\varepsilon_0}
\lim_{\omega \to 0} 
\frac{T}{\omega}
\int \frac{d^3q}{(2 \pi)^3} 
q^2 {\rm Im} \varepsilon_{\rm r} (q, \omega) .
\label{eq:general2}
\end{align}
In Eq.~(\ref{eq:general1}), 
the first and second terms
represent contributions from the magnetization and polarization  
currents, respectively. 
One may naively think that the contribution from the polarization current 
vanishes in the limit of $\omega_0\rightarrow 0$. 
However, this is not the case because of the singularity for $q \to 0$ 
as seen in the next section.

\section{Relaxation rates and the Knight shift for Dirac electrons} 
\label{S:5}

In this section, we apply the general expressions obtained in Sect.~\ref{S:4} 
to the Dirac electron system.
Substitution of Eqs.~(\ref{eq:main1}) and~(\ref{eq:main2}) for Dirac electrons
into Eqs.~(\ref{eq:general1}) and (\ref{eq:general2}) and some manipulations yield
\begin{align}
\left(\frac{1}{T_1}\right)_{\rm orb}
&= \frac{2 \pi}{3} e^2 \mu_0^2 \gamma_{\rm n}^2 {c^*}^4
T 
\int_{-\infty}^{\infty} d \epsilon 
\left[ - \frac{\partial f (\epsilon)}{\partial \epsilon} \right]
\frac{D^2(\epsilon)}{\epsilon^2} 
\log \frac{2(\epsilon^2 - \Delta^2)}{|\epsilon| \omega_0} ,
\label{eq:main_orb}
\\
\left(\frac{1}{T_1}\right)_{\rm Q}
&= \frac{7 \pi}{3} 
\frac{2 I + 3}{I^2 (2I - 1)} e^2 \mu_0^2 \gamma_{\rm n}^2 {c^*}^4
\frac{T}{U_0^2} 
\int_{-\infty}^{\infty} d \epsilon 
\left[ - \frac{\partial f (\epsilon)}{\partial \epsilon} \right]
\left( 1 + \frac{\Delta^2}{\epsilon^2} \right) D^2(\epsilon) ,
\label{eq:main_Q}
\end{align}
where $D(\epsilon)$ is the density of states of Dirac electrons as
\begin{align}
D(\epsilon) &= \int \frac{d^3k}{(2 \pi)^3} 
\left[ \delta \left( \epsilon - E_{\boldsymbol k} \right) 
+ \delta \left( \epsilon + E_{\boldsymbol k} \right) \right] 
\nonumber
\\
&= \frac{\epsilon \sqrt{\epsilon^2 - \Delta^2}}{2 \pi^2 {c^*}^3} 
\theta (\epsilon^2 - \Delta^2) .
\end{align}
In Eq.~(\ref{eq:main_Q}), we define a quantity $U_0$ with the dimension of energy 
(recovering $\hbar$) as
\begin{align}
U_0 \equiv 10 \, e \left( \frac{\hbar}{e^2} \right) \left( \frac{\hbar \gamma_{\rm n}}{Q} \right) 
\left( \frac{c^*}{c} \right)^2
= 2.075 \times 10^6 \, \frac{g}{Q/{\rm b}} \left( \frac{c^*}{c} \right)^2 
\, \mbox{[eV]},
\label{eq:U0}
\end{align}
where $g = \hbar \gamma_{\rm n}/\mu_{\rm N}$ is the g-factor of a nucleus 
with $\mu_{\rm N}$ being the nuclear magneton
and ${\rm b} = 10^{-28} {\rm m}^2$. 
It is to be noted that $(1/T_1)_{\rm orb}$ has 
contributions not only from the magnetization current but also 
the polarization current. 
To see this, 
we need to be careful with the terms proprtional to $\omega_0^2/q^4$ in
the integrand of Eq.~(\ref{eq:general1}) [see also Eqs.~(\ref{eq:main1}) and ~(\ref{eq:main2})]. 
The integrals of these terms with respect to $q$ 
do not vanish in the limit of $\omega_0\rightarrow 0$ 
and lead to  a cancellation in the nonlogarithmic term, 
leaving only the logarithmic term in Eq.~(\ref{eq:main_orb}). 

The full $T$ dependence of $\left( 1/T_1 \right)_{\rm orb}$ and $\left( 1/T_1 \right)_{\rm Q}$ 
are described by Eqs.~(\ref{eq:main_orb}) and~(\ref{eq:main_Q})  for Dirac electron systems.
Here we focus on the following two cases: 
\begin{enumerate}
\item[(i)]  For $|\mu| > \Delta = {m^*} {c^*}^2$ and $T / (|\mu| - \Delta) \to 0$, 
Eqs.~(\ref{eq:main_orb}) and~(\ref{eq:main_Q}) reduce to
\begin{align}
\left( \frac{1}{T_1T} \right)_{\rm orb} 
&= \frac{2 \pi}{3} \left( \frac{e \mu_0 \gamma_{\rm n}  k_{\rm F}}{2 \pi^2} \right)^2
\log \frac{2 |v_{\rm F}| k_{\rm F}}{\omega_0} , 
\label{eq:37}
\\
\left(\frac{1}{T_1T}\right)_{\rm Q}
&= 
\frac{7 \pi}{3}
\frac{2 I + 3 }{I^2 (2I - 1)} 
\left( \frac{e \mu_0 \gamma_{\rm n}  k_{\rm F}}{2 \pi^2} \right)^2
\left[  1 - \frac{1}{2} \left( \frac{v_{\rm F}}{c^*} \right)^2 \right] 
\left( \frac{{c^*}^2 k_{\rm F}}{v_{\rm F} U_0} \right)^2 ,
\label{eq:38}
\end{align}
where $k_{\rm F} = \sqrt{\mu^2 - \Delta^2}/{c^*}$ and 
$v_{\rm F} = {c^*} \sqrt{\mu^2 - \Delta^2}/\mu$ are 
the Fermi momentum and Fermi velocity, respectively. 
We note that there is a relationship of 
$\mu v_{\rm F} = {c^*}^2 k_{\rm F}$ between energy flux and momentum 
because Dirac electrons are ``relativistic'' particles. 
\item[(ii)] For $|\mu| \leq \Delta = {m^*} {c^*}^2$ and $\Delta /T \to 0$, 
Eqs.~(\ref{eq:main_orb}) and~(\ref{eq:main_Q}) reduce to
\begin{align}
\left( \frac{1}{T_1T} \right)_{\rm orb} 
&= \frac{1}{2 \pi} 
\left( \frac{e \mu_0 \gamma_{\rm n} T}{{3 c^*}} \right)^2
\left( 1.05 + \log \frac{2 T}{\omega_0} \right) ,
\label{eq:39}
\\
\left(\frac{1}{T_1T}\right)_{\rm Q}
&= 
\frac{49 \pi}{20} \frac{2 I + 3}{I^2 (2I - 1)} 
\left( \frac{e \mu_0 \gamma_{\rm n} T}{{3 c^*}} \right)^2 
\left( \frac{T}{U_0} \right)^2 .
\label{eq:40}
\end{align}
Thus, $\left( 1/T_1 \right)_{\rm orb}$ shows anomalous $T$ dependence as 
$\left( 1/T_1 \right)_{\rm orb} \sim T^3 \log (2T/\omega_0)$. 
On the other hand,    
$\left( 1/T_1 \right)_{\rm Q}$ is proportional to $T^5$ 
in the same way as the relaxation rate due to the Fermi contact interaction.  
\end{enumerate}
For Dirac electron systems such as bismuth, a typical value of ${c^*}/c$ is $10^{-3}$. 
Then, from Eq.~(\ref{eq:U0}), 
we estimate $U_0 \sim 10$ eV for most of the nuclei; 
for example, $U_0 \approx 55$ eV 
for the $^8$Li nucleus in the $\beta$-NMR experiment of 
Bi$_{0.9}$Sb$_{0.1}$~\cite{MacFarlane14}. 
Then, the last factors $( {c^*}^2 k_{\rm F}/v_{\rm F} U_0 )^2 = (\mu/U_0)^2$ and $(T/U_0)^2$ in Eqs.~(\ref{eq:38}) and~(\ref{eq:40}) 
are much smaller than $1$ for any $|\mu|$ and $T$ below the bandwidth cutoff $E_{\Lambda}$, 
which is on the order of $1$ eV. 
Thus, the relaxation rate $\left( 1/T_1 \right)_{\rm Q}$ due to the quadrupole interaction 
can be ignored compared with the relaxation rate $\left( 1/T_1 \right)_{\rm orb}$ due to 
the orbital interaction.  

Since we have shown $\left( 1/T_1 \right)_{\rm Q} \ll \left( 1/T_1 \right)_{\rm orb}$, 
we concentrate on $\left( 1/T_1 \right)_{\rm orb}$ in the following. 
The $T$ dependence of $\left( 1/T_1 \right)_{\rm orb}$ has been previously obtained 
numerically in Ref.~\cite{Hirosawa17}, where 
the right hand side of Eq.~(13) should be multiplied by $-2$ and 
numerical results of $1/T_1$ in Figs.~1--3 should be multiplied by $2$. 
A correct analytic expression of $\left( 1/T_1 \right)_{\rm orb}$ is given by Eq.~(\ref{eq:main_orb}) in this paper. 
Because the Dirac electron system reduces to the free-electron gas for $v_{\rm F}/{c^*} \ll 1$ 
with $\mu > \Delta$ while it is equivalent to two Weyl fermion systems for $\Delta = 0$, 
our result, Eq.~(\ref{eq:main_orb}), includes the results for 
the free-electron gas and Weyl fermion system, as seen below.
 
The relaxation rate $\left( 1/T_1 \right)_{\rm orb}$ due to the orbital interaction 
for the free-electron gas was obtained by Knigavko {\it et al.}~\cite{Knigavko07}. 
Their result (see Eq.~(19) in Ref.~~\cite{Knigavko07}) is described in our notation 
and SI units as
\begin{align}
\left( \frac{1}{T_1T} \right)_{\rm orb} 
&= \frac{2 \pi}{3} \left( \frac{e \mu_0 \gamma_{\rm n}  k_{\rm F}}{2 \pi^2} \right)^2
\left( \log \frac{4 \epsilon_{\rm F}}{\omega_0}  - 1 \right), 
\end{align}
where $\epsilon_{\rm F} = k_{\rm F}^2 /2 {m^*}$ is the Fermi energy. 
Thus, their result coincides with Eq.~(\ref{eq:37}) in the limit of $k_{\rm F} \to 0$ 
except for the nonlogarithmic term of $-1$. 
The absence of the nonlogarithmic term in Eq.~(\ref{eq:37}) is 
due to the contribution from the polarization current as mentioned before.

For the massless ($\Delta = 0$) case, 
from Eqs.~(\ref{eq:37}) and~(\ref{eq:39}), we find
\begin{align}
\left( \frac{1}{T_1} \right)_{\rm orb}  
&= \frac{e^2 \mu_0^2 \gamma_{\rm n}^2}{6 \pi^3 {c^*}^2}  T \times
\left\{ 
\begin{array}{lr}
\displaystyle{\mu^2 \log \frac{2 |\mu|}{\omega_0}}, 
&
|\mu| \gg T
\\
&
\\
\displaystyle{
T^2 \frac{\pi^2}{3} \log \frac{2 T}{\omega_0} ,
}
&
|\mu| \ll T
\end{array}
\right. ,
\label{eq:26}
\end{align}
where $|\mu| = {c^*}k_{\rm F}$ and the nonlogarithmic term of $1.05$ in Eq.~(\ref{eq:39}) 
is neglected.
On the other hand, the relaxation rate $\left( 1/T_1 \right)_{\rm orb}$ due to the orbital interaction for the Weyl fermion system was obtained by 
Okv\'{a}tovity {\it et al}.~\cite{Okvatovity16}.
By comparison between Eq.~(\ref{eq:26}) and their result (see Eq.~(16) in Ref.~\cite{Okvatovity16}),  
we confirm that 
$\left( 1/T_1 \right)_{\rm orb}$ in the massless Dirac electron system 
is twice as much as $\left( 1/T_1 \right)_{\rm orb}$ in the Weyl electron system 
(we suspect that the factor $52.7$ in Eq.~(16) of Ref.~\cite{Okvatovity16} 
may be equal to $(4 \pi)^2/3$).

Next, we discuss the Nnight shift $K_{\rm orb}$ due to the orbital interaction 
in Dirac electron systems. 
As shown in Sect.~\ref{S:4.2}, $K_{\rm orb}$ is equal to the orbital magnetic susceptibility 
$\chi_{\rm m}$. On the other hand, $\chi_{\rm m}$ for the Dirac electrons is given by 
Eqs.~(\ref{eq:diamag1}) and~(\ref{eq:diamag2}) in Sect.~\ref{S:2}. 
Thus, $K_{\rm orb}$ is obtained as
\begin{align}
K_{\rm orb}
&= - \frac{2 \alpha}{3 \pi } \frac{{c^*}}{c} 
\left[ \log \frac{E_{\Lambda}}{\Delta}
- \int_{\Delta}^{\infty} d \epsilon 
\frac{1 - f (-\epsilon) + f(\epsilon) }{\sqrt{\epsilon^2 - \Delta^2}}
\right] .
\label{eq:43}
\end{align}
As shown in Fig.~\ref{fig:1}, $K_{\rm orb} = \chi_{\rm m}$ is negative and 
its magnitude significantly increases for $|\mu| \lesssim \Delta$. 
For the massless case, in particular, $K_{\rm orb}$ is evaluated as 
\begin{align}
K_{\rm orb} = - 1.55 \times 10^{-3} \frac{c^*}{c} \log \frac{E_\Lambda}{2\, {\rm max}[|\mu|, T]}. 
\end{align}
Thus, for $\mu \approx 0$ and $T \to 0$, the orbital interaction gives rise to 
a large Knight shift.

It is emphasized that the large Knight shift due to the orbital interaction
is caused by the interband effect of a magnetic field. 
As seen from Eq.~(\ref{eq:general1}), 
on the other hand, the relaxation rate $1/T_1$ is determined only by the intraband effect. 
Therefore, the Korringa relation, which is satisfied in most metals as only the intraband excitations are allowed, is no longer valid for the Dirac electron systems.

\section{Discussion}
\label{S:6}

In the previous section, we obtained rigorous results of 
$\left( 1/T_1 T \right)_{\rm orb}$ for the 3D Dirac electron system 
and free-electron gas.
Two comments are in order with the obtained results. First, the zero-temperature result 
of $\left( 1/T_1T \right)_{\rm orb}$ for the Dirac electron system 
is formally equivalent to the zero-temperature result 
of $\left( 1/T_1T \right)_{\rm orb}$ for the free-electron gas, 
but $k_{\rm F}$ and $v_{\rm F}$ are identified as 
$k_{\rm F} = \sqrt{\mu^2 - \Delta^2}/{c^*}$ and 
$v_{\rm F} = {c^*} \sqrt{\mu^2 - \Delta^2}/\mu$, respectively [see Eq.~(\ref{eq:37})]. 
This feature is related to the fact that 
the Dirac electron system reduces to the free-electron gas for $k_{\rm F} \to 0$ 
with $\mu > \Delta$.
Second, the finite-temperature result 
of $\left( 1/T_1T \right)_{\rm orb}$ can be expressed as an integral 
of  the zero-temperature result of $\left( 1/T_1T \right)_{\rm orb}$ 
in the same manner as Eq.~(\ref{eq:diamag1}) for the orbital magnetic susceptibility 
[see Eq.~(\ref{eq:main_orb})]. This feature is valid for noninteracting systems, 
irrespective of dimensions~\cite{Maebashi17,Maldague78}. 
In this section, by noting these features, we give a prediction on 
the temperature dependence of 
$\left( 1/T_1T \right)_{\rm orb}$, which depends on the direction of an applied magnetic field, 
for quasi-2D Dirac electron systems.

Firstly, we rederive $\left( 1/T_1 T \right)_{\rm orb}$ for the 3D free-electron gas
by using the transverse conductivity 
$\sigma (q) = \lim_{\omega \to 0}{\rm Re} \sigma_{\rm T} (q,\omega)$ 
in the anomalous-skin-effect limit. 
For the free-electron gas, $\sigma (q)$ diverges as 
$\sigma (q) = e^2 k_{\rm F}^2/4\pi q$ for $q \to 0$~\cite{Glasser63}. 
Then, evaluating the integral over $q$ in 
$(1/T_1T)_{\rm orb} = \frac{4}{3}  \mu_0^2 \gamma_{\rm n}^2
 \int \frac{d^3q}{(2 \pi)^{3}} q^{-2} \sigma (q)$ 
[see Eq.~(\ref{eq:Worb})] by 
a lower cutoff at $q = \omega_0/|v_{\rm F}|$ and an upper cutoff $q = 2 k_{\rm F}$, 
we obtain Eq.~(\ref{eq:37}) again.

For the quasi-2D free-electron gas (i.e., metallic layers where electrons in each layer is described by the 2D free-electron gas), by using the same method as shown above, 
Lee and Nagaosa obtained the relaxation rates $(1/T_{1\parallel})_{\rm orb}$ and $(1/T_{1\perp})_{\rm orb}$ due to the orbital interaction 
when the magnetic field is applied parallel and perpendicular to the layers, 
respectively~\cite{Lee91}. 
With use of our cutoff scheme for the $q$ integral, 
their result in SI units is given by
\begin{align} 
\left(\frac{1}{T_{1\parallel}T}\right)_{\rm orb} 
= \frac{3}{2} \left(\frac{1}{T_{1\perp}T}\right)_{\rm orb} 
&=   
\frac{\mu_0^2 \gamma_{\rm n}^2}{4 d} \int \frac{d^2q }{(2 \pi)^2} q^{-1}\sigma (q)
= \left( \frac{\gamma_{\rm n} \mu_0 e}{4 \pi} \right)^2  \frac{k_{\rm F}}{d} \log 
\frac{2 |v_{\rm F}| k_{\rm F}}{\omega_0}, 
\label{eq:45}
\end{align}
where $d$ is a distance between nearest neighbor layers and 
$\sigma (q) = e^2 k_{\rm F}/2\pi q$ is the anomalous-skin-effect expression 
of the transverse conductivity in 2D. 
We assume that Eq.~(\ref{eq:45}) is valid for the quasi-2D Dirac electron system 
in the limit of zero temperature
by identifying $k_{\rm F} = \sqrt{\mu^2 - \Delta^2}/{c^*}$ and 
$v_{\rm F} = {c^*} \sqrt{\mu^2 - \Delta^2}/\mu$ 
with $|\mu| > \Delta$. 
Then, the finite-temperature results 
of $\left( 1/T_{1\parallel}T\right)_{\rm orb}$ and $\left( 1/T_{1\perp}T\right)_{\rm orb}$ 
are derived from integrals of their zero-temperature results with respect to 
the chemical potential as  
\begin{align}
\left(\frac{1}{T_{1\parallel}T}\right)_{\rm orb} 
= \frac{3}{2} \left(\frac{1}{T_{1\perp}T}\right)_{\rm orb} 
= \left( \frac{\gamma_{\rm n} \mu_0 e}{4 \pi} \right)^2  
\int d \epsilon 
\left[ - \frac{\partial f (\epsilon)}{\partial \epsilon} \right] 
\frac{\sqrt{\epsilon^2 - \Delta^2}}{{c^*}d}
\log \frac{2 (\epsilon^2 - \Delta^2)
}{|\epsilon| \omega_0}, 
\label{eq:46}
\end{align}
where the integral over $\epsilon$ is restricted by $|\epsilon| > \Delta$.

For the massless case, 
Eq.~(\ref{eq:46}) reduces to  
$(1/T_{1\parallel})_{\rm orb} = (3/2) (T_{1\perp})_{\rm orb} \sim T E \log (2 E/\omega_0)$, 
where $E$ is the maximum of $T$ and $|\mu|$. 
We note that the power of $E$ corresponds to the power of $k_{\rm F}$ in 
the transverse conductivity $\sigma(q)$. 
For the quasi-2D massless Dirac electron system with $\mu = 0$, 
in particular, these relaxation rates due to the orbital interaction 
show the $T^2 \log (2T/\omega_0)$ dependence, 
which overcomes the $T^3$ dependence from the Fermi contact interaction in $1/T_1$ 
for $T \to 0$.

\section{Concluding Remarks}
\label{S:7}

We have investigated the nuclear magnetic relaxation rate and Knight shift 
in the presence of the orbital and quadrupole interactions for 
3D Dirac electron systems. 
We have derived general expressions of Eqs.~(\ref{eq:general1}) and~(\ref{eq:general2}) 
for the relaxation rates $\left( 1/T_1 \right)_{\rm orb}$ and $\left( 1/T_1 \right)_{\rm Q}$ 
due to the orbital and quadrupole interactions, respectively, 
which are written in terms of the dynamic magnetic susceptibility $\chi_{\rm m}(q,\omega)$ 
and relative permittibity $\varepsilon_{\rm r} (q, \omega)$. 
In particular, the expression of $\left( 1/T_1 \right)_{\rm orb}$ includes 
contributions not only from the magnetization current but also from 
the polarization current. 
By using the results of $\chi_{\rm m}(q,\omega)$ and $\varepsilon_{\rm r} (q, \omega)$ 
in Ref.~\cite{Maebashi17},
we have obtained rigorous expressions of 
Eqs.~(\ref{eq:main_orb}) and~(\ref{eq:main_Q}) for $\left( 1/T_1 \right)_{\rm orb}$ and 
$\left( 1/T_1 \right)_{\rm Q}$,   
and shown that $\left( 1/T_1 \right)_{\rm Q}$ is much smaller than 
$\left( 1/T_1 \right)_{\rm orb}$ for the Dirac electron systems. 

The relaxation rate $\left( 1/T_1 \right)_{\rm orb}$ due to the orbital interaction 
diverges logarithmically for the nuclear Larmor frequency $\omega_0 \to 0$, 
and anomalously depends on temperature $T$ and chemical potential $\mu$.
When $\mu$ lies in the conduction or valence bands, 
we have obtained 
$\left( 1/T_1 \right)_{\rm orb} \propto T k_{\rm F}^2 \log (2 |v_{\rm F}| k_{\rm F}/\omega_0)$ 
for $T \to 0$, 
where $k_{\rm F}$ and $v_{\rm F}$ are the Fermi momentum and Fermi velocity, respectively 
[see Eq.~(\ref{eq:37})]. 
When $\mu$ is inside the band gap, 
we have obtained 
$\left( 1/T_1 \right)_{\rm orb} \sim T ^3 \log (2 T/\omega_0)$ 
for $T$ above the band gap [see Eq.~(\ref{eq:39})]. 
These results for the Dirac electrons are consistent with 
the previous results for the free-electron gas  in Ref.~~\cite{Knigavko07} 
and Weyl fermions in Ref.~\cite{Okvatovity16}, 
although our results include the correct nonlogarithmic term by taking account of 
a contribution from the polarization current.

The Knight shift $K_{\rm orb}$ due to the orbital interaction also shows 
anomalous dependences on $T$ and $\mu$ for Dirac electron systems. 
The Korringa relation does not hold between $K_{\rm orb}$ and 
$\left( 1/T_1 \right)_{\rm orb}$ because the anomalous dependences in $K_{\rm orb}$ 
are caused by the interband particle-hole excitations across the small band gap  
as given by  Eq.~(\ref{eq:43}) 
while $\left( 1/T_1 \right)_{\rm orb}$ is governed by the intraband excitations.  

It is to be noted that the large diamagnetism in $\alpha$-(BEDT-TTF)$_2$I$_3$, 
which is not a spin-orbit coupled system but a quasi-2D massless Dirac electron system, 
was also theoretically predicted
under a magnetic field perpendicular to the conducting layers~\cite{Fukuyama07,Kobayashi08}. 
Thus, the large Knight shift due to the orbital interaction is expected for the perpendicular 
magnetic field. For the quasi-2D massless Dirac electron system with $\mu = 0$, 
on the other hand, we have given a prediction on the $T$ dependences of the relaxation rates 
$(1/T_{1\parallel})_{\rm orb}$ and $(T_{1\perp})_{\rm orb}$ for the parallel and perpendicular 
fields, respectively, as  
$(1/T_{1\parallel})_{\rm orb} = (3/2) (T_{1\perp})_{\rm orb} \sim T^2 \log (2 T/\omega_0)$.
Recently, NMR experiments have been carried out in $\alpha$-(BEDT-TTF)$_2$I$_3$  
with a magnetic field parallel to the conducting layers for low 
temperatures~\cite{Hirata11,Hirata17}. 
It would be very interesting to see how they change with a perpendicular magnetic field.

\vspace{0.5cm}
\noindent
\textbf{Acknowledgments} 

We would like to thank Y. Fuseya, K. Kanoda, H. Matsuura, and N. Okuma 
for fruitful discussions and comments.
This work was supported by the Japan Society for the Promotion of Science through 
Program for Leading Graduate Schools (MERIT) and  
a Grant-in-Aid for Scientific Research on ``Multiferroics in Dirac electron materials'' (No.15H02108).


\begin{thebibliography}{00}

\bibitem{Wolff64} P. A. Wolff, J. Phys. Chem. Solids \textbf{25}, 1057 (1964).
\bibitem{Fuseya15} For a review, see Y. Fuseya, M. Ogata, and H. Fukuyama, J. Phys. Soc. Jpn. \textbf{84}, 012001 (2015).
\bibitem{Shoenberg36} D. Shoenberg and M. Z. Uddin, Proc. R. Soc. Lond. Ser. A \textbf{156}, 687 (1936).
\bibitem{Wehrli68} L. Wehrli, Phys. Kondens. Mater.  \textbf{8}, 87 (1968).
\bibitem{Boyle60} W. S. Boyle and A. D. Brailsford, Phys. Rev. \textbf{120}, 1943 (1960).
\bibitem{Edelman75} V. S. \`Edel'man, Zh. Eksp. Theor. Fiz. \textbf{68}, 257 (1975) [Sov. Phys.-JETP, \textbf{41}, 235 (1975) ]. 
\bibitem{Maebashi17} H. Maebashi, M. Ogata, and H. Fukuyama, J. Phys. Soc. Jpn. \textbf{86}, 083702 (2017).
\bibitem{MacFarlane14} W. A. MacFarlane, C. B. L. Tschense, T. Buck, K. H. Chow, D. L. Cortie, A. N. Hariwal,
R. F. Kiefl, D. Koumoulis, C. D. P. Levy, I. McKenzie, F. H. McGee, G. D. Morris, M. R.
Pearson, Q. Song, D. Wang, Y. S. Hor, and R. J. Cava, Phys. Rev. B \textbf{90}, 214422 (2014).
\bibitem{Okvatovity16} Z. Okv\'{a}tovity, F. Simon, and B. D\'{o}ra, Phys. Rev. B \textbf{94}, 245141 (2016).
\bibitem{Yasuoka16} H. Yasuoka, T. Kubo, Y. Kishimoto, D. Kasinathan, M. Schmidt, B. Yan, Y. Zhang, H. Tou, C. Felser, A. P. Mackenzie, and M. Baenitz, Phys. Rev. Lett. \textbf{118}, 236403 (2017).
\bibitem{Hirosawa17} T. Hirosawa, H. Maebashi, and M. Ogata, J. Phys. Soc. Jpn. \textbf{86}, 063705 (2017).
\bibitem{Luttinger55} J. M. Luttinger and W. Kohn, Phys. Rev. \textbf{97}, 869 (1955). 
\bibitem{Cohen60} M. H. Cohen and E. I. Blount, Phil. Mag. \textbf{5}, 115 (1960).
\bibitem{Fukuyama70} H. Fukuyama and R. Kubo, J. Phys. Soc. Jpn. \textbf{28}, 570 (1970).
\bibitem{Fukuyama71} H. Fukuyama, Prog. Theor. Phys. \textbf{45}, 704 (1971). 
\bibitem{Ogata15} M. Ogata and H. Fukuyama, J. Phys. Soc. Jpn. \textbf{84}, 124708 (2015).
\bibitem{Ogata16-1} M. Ogata, J. Phys. Soc. Jpn. \textbf{85}, 064709 (2016).
\bibitem{Ogata16-2} M. Ogata, J. Phys. Soc. Jpn. \textbf{85}, 104708 (2016).
\bibitem{Ogata17} M. Ogata, J. Phys. Soc. Jpn. \textbf{86}, 044713 (2017).
\bibitem{Abragam61} A. Abragam, {\it Principles of Nuclear Magnetism} (Clarendon Press, Oxford, UK, 1961).
\bibitem{Obata63} Y. Obata, J. Phys. Soc. Jpn. \textbf{18}, 1020 (1963).
\bibitem{Obata64} Y. Obata, J. Phys. Soc. Jpn. \textbf{19}, 2348 (1964).
\bibitem{Maldague78} P. F. Maldague, Surf. Sci. \textbf{73}, 296 (1978).
\bibitem{Fuseya14} Y. Fuseya, M. Ogata, and H. Fukuyama, J. Phys. Soc. Jpn. {\bf 83}, 074702 (2014).
\bibitem{Kubo57} R. Kubo, J. Phys. Soc. Jpn. \textbf{12}, 570 (1957).
\bibitem{Jackson99} J. D. Jackson,  {\it Classical electrodynamics} (Wiley, Hoboken, NJ, 1999). 
\bibitem{Lee91} P. Lee and N. Nagaosa, Phys. Rev. B \textbf{43}, 1223 (1991).
\bibitem{Chklovskii92} D. B. Chklovskii and P. Lee, Phys. Rev. B \textbf{45}, 5240 (1992).
\bibitem{Knigavko07} A. Knigavko, B. Mitrovi\'{c}, and K. V. Samokhin, Phys. Rev. B \textbf{75}, 134506 (2007).
\bibitem{Glasser63} M. L. Glasser, Phys. Rev. \textbf{129}, 472 (1963).
\bibitem{Fukuyama07} H. Fukuyama, J. Phys. Soc. Jpn. \textbf{76}, 043711 (2007).
\bibitem{Kobayashi08} A. Kobayashi, Y. Suzumura, and H. Fukuyama, J. Phys. Soc. Jpn. \textbf{77}, 064718 (2008).
\bibitem{Hirata11} M. Hirata, K. Ishikawa, K. Miyagawa, K. Kanoda, and M. Tamura, Phys. Rev. B 
\textbf{84}, 125133 (2011).
\bibitem{Hirata17} M. Hirata, K. Ishikawa, G. Matsuno, A. Kobayashi, K. Miyagawa, M. Tamura, C. Berthier, and K. Kanoda, arXiv:1702.00097. 

\end{thebibliography}
\end{document}